\documentclass{article}
\pdfoutput=1

\usepackage{arxiv}

\usepackage[utf8]{inputenc} 
\usepackage[T1]{fontenc}    
\usepackage{hyperref}       
\usepackage{url}            
\usepackage{booktabs}       
\usepackage{amsfonts}  
\usepackage{amsthm}
\usepackage{nicefrac}       
\usepackage{microtype}      
\usepackage{lipsum}

\usepackage[nocompress]{cite}
\newtheorem{definition}{Definition}

\newtheorem{proposition}{Proposition}

\usepackage{times}
\usepackage{booktabs} 
\usepackage{paralist}
\usepackage{multirow}
\newcommand{\cut}[1]{}
\usepackage{graphicx}
\usepackage{booktabs}
\usepackage{color}
\usepackage{amsmath}
\usepackage{amssymb}

\usepackage{url}
\usepackage{wrapfig}
\usepackage{algpseudocode}
\usepackage{algorithm}
\usepackage{epstopdf}
\usepackage{multirow}
\usepackage{rotating}
\usepackage{relsize}
\usepackage{appendix}
\usepackage{bbm}
\usepackage{mathtools}

\title{Detecting local community structures in Social Networks Using Concept Interestingness}

\author{
  Mohamed-Hamza Ibrahim\thanks{Corresponding author} \\
  University of Quebec in Outaouais \\
  Quebec, Canada \\
  \texttt{ibrm05@uqo.ca} \\
   \And
 Rokia Missaoui \\
  University of Quebec in Outaouais \\
  Quebec, Canada \\
  \texttt{rokia.missaoui@uqo.ca} \\
  \And
 Abir Messaoudi \\
  University of Quebec in Outaouais \\
  Quebec, Canada \\
  \texttt{mesa08@uqo.ca} \\
}

\begin{document}
\maketitle
\begin{abstract}
One key challenge in Social Network Analysis is to design an efficient and accurate community detection procedure as a means to discover intrinsic structures and extract relevant information. In this paper, we introduce a novel strategy called (\textbf{COIN}), which exploits \underline{\textbf{CO}}ncept \underline{\textbf{IN}}terestingness measures to detect communities based on the concept lattice construction of the network. Thus, unlike off-the-shelf community detection algorithms, COIN leverages relevant conceptual characteristics inherited from Formal Concept Analysis to discover substantial local structures. On the first stage of COIN, we extract the formal concepts that capture all
the cliques and bridges in the social network. On the second stage, we use the stability index to remove noisy bridges between communities and then percolate (merge)
relevant adjacent cliques. Our experiments on several real-world social
networks show that COIN can quickly detect communities more
accurately than existing prominent algorithms such as Edge betweenness, \cut{Louvain, Walktrap, }Fast greedy modularity, and Infomap.
\end{abstract}

\keywords{Social network analysis, community detection, Formal
Concept Analysis, formal concept interestingness.}

\maketitle

\section{Introduction}
\label{Intro}

Given a social network $\mathcal{U} = (\mathcal{G},\mathcal{I})$, where the node set $\mathcal{G}$ includes
the objects in the social network, and the edge set $\mathcal{I} =\{(g_i,g_j) \mid g_i,g_j \in \mathcal{G} \}$ denotes the relationship between objects. The problem is to detect all possible communities by dividing the network into groups of nodes (i.e., meaningful connected components) based on the hidden relevant relationships in $\mathcal{U}$. 

The discovery of cohesive groups, cliques and
communities inside a network is one of the most studied topics in
social network analysis. It has attracted many researchers in
sociology, biology, computer science, physics, criminology, and so
on. Community detection \cite{Fortunato2010,Fortunato2016} aims at finding clusters as sub-graphs within
a given network. A community is then a cluster where many edges connect
nodes of the same group and few edges link nodes of different
groups. For instance, a community in the social network \textit{LinkedIn} may represent members with a similar professional profile.

Community detection algorithms can be mainly categorized into two groups \cite{Fortunato2010}: (i) Agglomerative procedures, in which nodes/groups are iteratively
merged if they are similar, and (ii) divisive algorithms, in which clusters are iteratively decomposed by cutting the edges between less similar vertices.
A finer categorization of community detection algorithms includes the following main kinds: hierarchical clustering, modularity maximization, clique (and variants such as $n$-clique and $k$-plex) identification, block-modeling, and spectral graph partitioning \cite{girvan2002community1,newman2004finding,Fortunato2010}.

In this paper, we leverage Formal Concept Analysis (FCA) and the stability index of identical concepts to find relevant cliques and irrelevant bridges. Formal Concept Analysis is a mathematical formalism for data analysis \cite{Ganter+1999} that uses a formal context as input to construct a set of formal concepts organized in a concept lattice. 
It has been successfully used in several areas of computer science to discover patterns such as homogeneous groups or association rules. In \cite{Freeman1996}, Freeman was the first to use FCA for community detection in one-mode data social networks. His FCA-based method starts with an adjacency matrix where objects are
individuals and attributes are maximal cliques of a size at least equal to $3$, constructs the concept lattice, identifies and then eliminates special cliques and edges to finally get the communities. Indeed, this opens the door to a promising research area of using cliques to detect communities in the social network. In this context, several clique-based methods have been introduced in the literature, including clique percolation methods \cite{palla2005uncovering,evans2010clique,adamcsek2006cfinder}. More recently, Hao et. al. \cite{hao2017k} define a new method that identifies $k$-equiconcepts to further generate $k$-cliques in social networks. Our COIN method can be seen as akin to these clique-based methods. But a basic aspect of COIN is the use of Formal Concept Analysis theory to better understand the network topology and the exploitation of concept interestingness measure such as stability index to discover communities by identifying relevant and irrelevant parts of the social network.

The paper is organized as follows. Section \ref{Back} gives a background about social network analysis and FCA while Section \ref{Coin} describes our method for community detection in \textit{one-mode data networks} using FCA and interestingness measures.
In Section \ref{Tests} we provide an empirical of our method against existing ones. We finally conclude the paper and describe further work in Section \ref{Conc}.

\section{Background}
\label{Back}
\begin{figure}[!htbp]
\centering
    \includegraphics[width=75mm]{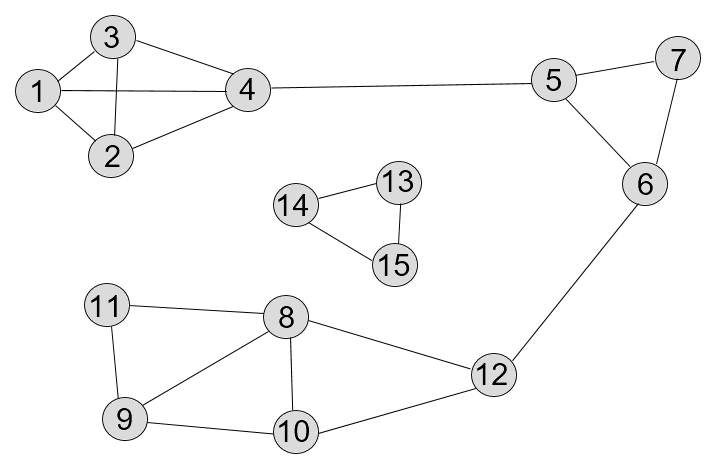}  
      \caption{Undirected graph $\mathcal{U}$ as an excerpt of the LinkedIn network.}
      \label{toygraph}
\end{figure}

This section will briefly review the main concepts that support the comprehension of our COIN community detection method by using an illsutrative example, which is an excerpt of a LinkedIn connection network and contains $15$ members of the LARIM team at University of Quebec in Outaouais. As shown in Figure~\ref{toygraph}, the network is modeled as an undirected graph $\mathcal{U} = (\mathcal{G},\mathcal{I})$, where $\mathcal{G}$ is a set of $15$ nodes representing members, and $\mathcal{I}$ is a set of edges where an edge $(g_i,g_j) \in \mathcal{I}$ connects two members, $g_i,g_j \in \mathcal{G}$, if they have a link on LinkedIn. Let us now express our basic notation.

\subsection{Basic Notation and Definitions}
\begin{definition}[Clique] Let $\mathcal{U} = (\mathcal{G},\mathcal{I})$ be an undirected graph defined over the objects $\mathcal{G}$. A clique of size $k$ in $\mathcal{U}$ is a subset $Q \subset \mathcal{G}$ such that for any two nodes (i.e., objects) $g_i$ and $g_j \in Q$, there exists an edge  (i.e., a binary relation) $(g_i,g_j) \in \mathcal{I}$.
\label{Cl}
\end{definition}

In the sequel, we will express a clique by a set of nodes without reference to the edges. For instance, the set $Q=\{5,6,7\}$ represents a clique of size $3$. 

\begin{definition}[Maximal clique] A clique $Q \subset \mathcal{G}$ is \textbf{maximal} if it cannot be extended to include one more adjacent object node.
\label{MC}
\end{definition}
For example, $Q=\{1,2,3,4\}$ is a maximal clique of size $4$.  

\begin{definition}[Isolated maximal clique] A maximal clique $Q \subset \mathcal{G}$ is \textbf{isolated} if $\forall g_i \in Q$ and $\forall g_j \in \mathcal{G}-Q$ we have $(g_i,g_{j}) \notin \mathcal{I}$. That is, there is no edge that connects an object in the maximum clique $Q$ to any object outside it.
\label{IMC}
\end{definition}
 For example, $Q=\{13,14,15\}$ represents an isolated maximal clique in Figure~\ref{toygraph}.   

\begin{definition}[Bridge or Cut-edge] An edge $(g_i,g_{j}) \in \mathcal{I}$ is a bridge iff it is not contained in any cycle and its removal increases the number of connected components in the graph $\mathcal{U}$.  
\end{definition}

\begin{definition}[Non-trivial Bridge] A Bridge $(g_i,g_{j}) \in \mathcal{I}$ is \textbf{non-trivial} iff its end vertices $g_i$ and $g_{j}$ have a degree (\emph{i.e.}, number of neighbors) greater than $2$.  
\end{definition}

For example, 
the edge $(4,5)$ 
is a non-trivial bridge since the end vertices $g_4$ and $g_5$ have a degree equal to $4$ and $3$, respectively.

\subsection{Formal Concept Analysis}
In the following we recall key notions of FCA that will be used in this paper.

\begin{definition}[Formal
context] It is a triple $\mathbb{K} = (\mathcal{G},\mathcal{M},\mathcal{I})$, where $\mathcal{G}$ is a set of objects, $\mathcal{M}$ a set of attributes, and $\mathcal{I}$ a binary relation between $\mathcal{G}$ and $\mathcal{M}$ with $\mathcal{I} \subseteq \mathcal{G} \times \mathcal{M}$. For $g \in \mathcal{G}$ and $m \in \mathcal{M}, (g,m) \in \mathcal{I}$ holds (i.e., $(g,m)=1$) iff the object $g$ has the attribute $m$, and otherwise $(g,m) \notin \mathcal{I}$ (i.e., $(g,m)=0$).  
\label{defFC}
\end{definition}
 Given arbitrary subsets $A \subseteq \mathcal{G}$ and $B \subseteq \mathcal{M}$, the following derivation operators are defined:
\begin{equation*}
   A^{\prime{}} = \{m \in \mathcal{M} \mid \forall g \in A, (g,m) \in \mathcal{I} \}, \; A \subseteq \mathcal{G}  
\end{equation*}
\begin{equation*}
   B^{\prime{}} = \{g \in \mathcal{G} \mid \forall m \in B, (g,m) \in \mathcal{I} \}, \; B \subseteq \mathcal{M} 
\end{equation*}
where $A^{\prime{}}$ is the set of attributes common to all objects of $A$ and $B^{\prime{}}$ is the set of objects sharing all attributes from $B$. The closure operator $(.)^{\prime{}\prime{}}$ implies the double application of $(.)^{\prime{}}$, which is extensive, idempotent and monotone. The subsets $A$ and $B$ are closed when $A=A^{\prime{}\prime{}}$, and $B=B^{\prime{}\prime{}}$. 
\begin{definition}[Formal concept]
The pair $c=(A,B)$ is called a \textit{formal concept} of $\mathbb{K}$ with \textit{extent} $A$ and \textit{intent} $B$ if both $A$ and $B$ are closed and $A^{\prime{}}=B$, and $B^{\prime{}}=A$. 
\end{definition}
For a finite intent (or extent) set of $w$ elements, we use $\mathcal{P}(.)$ to denote its power set with a number of subsets equal to $n=2^w$, i.e., the set of all its subsets, including the empty set and the set itself.

\begin{definition}[Partial order relation $\leq$]
A concept $c_1=(A_1,B_1)$ $\leq$ $c_2=(A_2,B_2)$ if:
\begin{equation}
A_1 \subseteq A_2 \iff  B_1 \supseteq B_2    
\end{equation} 
\end{definition}
In this case, $c_2$ is called a superconcept (or upper neighbor or successor) of $c_1$, and $c_1$ is called a subconcept (or lower neighbor or predecessor) of $c_2$. The set of all concepts of the formal context $\mathbb{K}$ is expressed by $\mathcal{C}(\mathbb{K})$ or simply $\mathcal{C}$.

\begin{definition}[Concept Lattice]
The concept lattice of a formal context $\mathbb{K}$, denoted by $\mathcal{L}(\mathbb{K})=(\mathcal{C},\leq)$, is a Hasse graphical diagram that represents all formal concepts $\mathcal{C}$ together with the partial order that holds between them. In $\mathcal{L}(\mathbb{K})$, each node represents a concept with its extent and its intent while the edges represent the partial order between concepts. 
\end{definition}
There are several methods (cf. \cite{Ganter+1999,nourine1999fast,lindig2000fast,valtchev2002partition,choi2009faster}) that build the lattice, \emph{i.e.},
compute all the concepts together with the partial order. 

One-mode data networks contain only one type of nodes and relations. Hence, we can simply adapt the formal context (in Definition~\ref{defFC}) to define a one-mode data context as follows.

\begin{definition}[One-mode formal
context] It is a formal context $\Tilde{\mathbb{K}} = (\mathcal{G},\mathcal{G},\mathcal{I})$ in which the two sets of objects and attributes are identical, i.e., $\mathcal{G} \equiv \mathcal{M}$, and $\mathcal{I}$ is a set of relations defined on $\mathcal{G}$ with $\mathcal{I} \subseteq \mathcal{G} \times \mathcal{G}$. For $g_i,g_j \in \mathcal{G}$, $(g_i,g_j) \in \mathcal{I}$ holds iff object $g_i$ is linked to $g_j$ or $g_i = g_j$. \end{definition}

\subsection{Concept interestingness}
Interestingness (quality) measures of a formal concept $c=(A, B)$ are commonly used to assess its \textit{relevancy}. While several interestingness measures have been introduced to select relevant concepts \cite{buzmakov2014concept,kuznetsov2007stability,roth2008succinct,klimushkin2010approaches}, the stability \cut{and separation indices} index of $c$, $\sigma(c)$ \cut{and $\alpha(c)$, have} has been found to be the most prominent for selecting relevant concepts \cite{Kuznetsov2015}. 

\begin{definition}[Stability Index] Let $\mathbb{K} = (\mathcal{G},\mathcal{M},\mathcal{I})$ be a formal context and $c=(A,B)$ a formal
concept of $\mathbb{K}$. The \textit{intensional stability} $\sigma(c)$ can be computed as \cite{kuznetsov2007stability,Babin2012,buzmakov2014concept}:
\begin{equation}\label{stability1}
    \sigma(c) = \frac{\mid \{e \in \mathcal{P}(A) | e^{\prime{}}= B\}\mid}{2^{|A|}}
\end{equation}
\end{definition}

In Equation~\eqref{stability1}, intensional stability $\sigma(c)$ measures the strength of dependency between the intent $B$ and the objects of the extent $A$. More precisely, it expresses the probability to maintain $B$ closed when a subset of noisy objects in $A$ are deleted with equal probability. In fact, this measure quantifies the amount of noise in the extent $A$ and overfitting in the intent $B$. The numerator of $\sigma(c)$ in Eq.~\eqref{stability1} can be computed exactly by identifying and counting the minimal generators of the concept \cite{roth2008succinct,zhi2014calculation}. Such computation takes a time complexity of $O(L^2)$ \cite{roth2008succinct,zhi2014calculation}, where $L$ is the size of the concept lattice, and requires the lattice construction which needs a time complexity of $O(|\mathcal{G}|^2 \cdot |\mathcal{M}| \cdot L)$ \cite{muangprathub2014novel,kuznetsov1999learning}. However, we have recently designed an efficient method to approximate the stability index using the low-discrepancy sampling (LDS) approach \cite{IbrahimMRokia18}. Taking a set $S$ of uniformly distributed samples from the intent powerset of a concept $c$, the LDS method needs a time complexity $\xi$ of $O(|S|)$ to estimate the stability index of $c$ with a convergence rate (i.e., sampling error) equals to $O\big(\frac{\log |S|}{|S|}\big)$.  
\section{COIN for Community Detection}
\label{Coin}
At a conceptual level, our overall COIN method consists of the following key elements. First, we build the formal context and construct the concept lattice of the social network. Second, we extract from the lattice the whole set of concepts that represent all the cliques and bridges in the social graph. Third, we use a concept interestingness measure to identify noisy bridges and relevant cliques. Finally, we remove the most noisy bridges and percolate (merge) the adjacent relevant cliques to detect communities. 

\subsection{Building a Formal Context for a Social Network}
In COIN, the first task is to build the one-mode formal context of the social network $\mathcal{U} = (\mathcal{G},\mathcal{I})$ by computing the symmetrical \textit{modified adjacency matrix} \cite{hao2017k} as follows:
\begin{equation}\label{Fcontext}
\begin{array}{lll}
\Tilde{\mathbb{K}}(\mathcal{G},\mathcal{G},\mathcal{I}) =\begin{cases}
(g_i,g_j) = 1 & \text{If $\exists~(g_i,g_j) \in \mathcal{I}, \; i \neq j$}\\
(g_i,g_j) = 1 & \text{if $\; i = j$} \\
(g_i,g_j) = 0 & \text{Otherwise}.
 \end{cases}
 \end{array}
\end{equation}
In Eq.~\eqref{Fcontext}, we assign $0$ to the element of $\Tilde{\mathbb{K}}$ in the row $i$ and column $j$ if the object (node) $g_i$ is not connected to the object $g_j$ in the graph $\mathcal{U}$. Otherwise, we assign $1$ to it. Note that the diagonal elements are assigned the value $1$. 

For example, the constructed formal context $\Tilde{\mathbb{K}}$ of our example is represented in Table~\ref{toycontext}. 

\begin{table}[!htbp]
\centering
\caption{The formal context $\Tilde{\mathbb{K}}$ for the network of Figure~\ref{toygraph}.}
\label{my-label}
\tabcolsep 3.5pt
\begin{tabular}{|l|lllllllllllllll|}
\hline
\textbf{$\mathcal{G}$} & \textbf{1} & \textbf{2} & \textbf{3} & \textbf{4} & \textbf{5} & \textbf{6} & \textbf{7} & \textbf{8} & \textbf{9} & \textbf{10} & \textbf{11} & \textbf{12} & \textbf{13} & \textbf{14} & \textbf{15} \\
\hline
\textbf{1} & 1 & 1 & 1 & 1 & 0 & 0 & 0 & 0 & 0 & 0 & 0 & 0 & 0 & 0 & 0 \\
\textbf{2} & 1 & 1 & 1 & 1 & 0 & 0 & 0 & 0 & 0 & 0 & 0 & 0 & 0 & 0 & 0 \\
\textbf{3} & 1 & 1 & 1 & 1 & 0 & 0 & 0 & 0 & 0 & 0 & 0 & 0 & 0 & 0 & 0 \\
\textbf{4} & 1 & 1 & 1 & 1 & 1 & 0 & 0 & 0 & 0 & 0 & 0 & 0 & 0 & 0 & 0 \\
\textbf{5} & 0 & 0 & 0 & 1 & 1 & 1 & 1 & 0 & 0 & 0 & 0 & 0 & 0 & 0 & 0 \\
\textbf{6} & 0 & 0 & 0 & 0 & 1 & 1 & 1 & 0 & 0 & 0 & 0 & 1 & 0 & 0 & 0 \\
\textbf{7} & 0 & 0 & 0 & 0 & 1 & 1 & 1 & 0 & 0 & 0 & 0 & 0 & 0 & 0 & 0 \\
\textbf{8} & 0 & 0 & 0 & 0 & 0 & 0 & 0 & 1 & 1 & 1 & 1 & 1 & 0 & 0 & 0 \\
\textbf{9} & 0 & 0 & 0 & 0 & 0 & 0 & 0 & 1 & 1 & 1 & 1 & 0 & 0 & 0 & 0 \\
\textbf{10} & 0 & 0 & 0 & 0 & 0 & 0 & 0 & 1 & 1 & 1 & 0 & 1 & 0 & 0 & 0 \\
\textbf{11} & 0 & 0 & 0 & 0 & 0 & 0 & 0 & 1 & 1 & 0 & 1 & 0 & 0 & 0 & 0 \\
\textbf{12} & 0 & 0 & 0 & 0 & 0 & 1 & 0 & 1 & 0 & 1 & 0 & 1 & 0 & 0 & 0 \\ \cut{\cline{14-16}}
\textbf{13} & 0 & 0 & 0 & 0 & 0 & 0 & 0 & 0 & 0 & 0 & 0 & 0 & \cut{\multicolumn{1}{|l}{1}}1 & 1 & 1 \\
\textbf{14} & 0 & 0 & 0 & 0 & 0 & 0 & 0 & 0 & 0 & 0 & 0 & 0 & \cut{\multicolumn{1}{|l}{1}}1 & 1 & 1 \\
\textbf{15} & 0 & 0 & 0 & 0 & 0 & 0 & 0 & 0 & 0 & 0 & 0 & 0 & \cut{\multicolumn{1}{|l}{1}}1 & 1 & 1 \\
\cut{\cline{14-16}}
\hline
\end{tabular}\label{toycontext}
\end{table}

\begin{figure}[!htbp]
\centering
    \includegraphics[width=3.9in,height=2.8in]{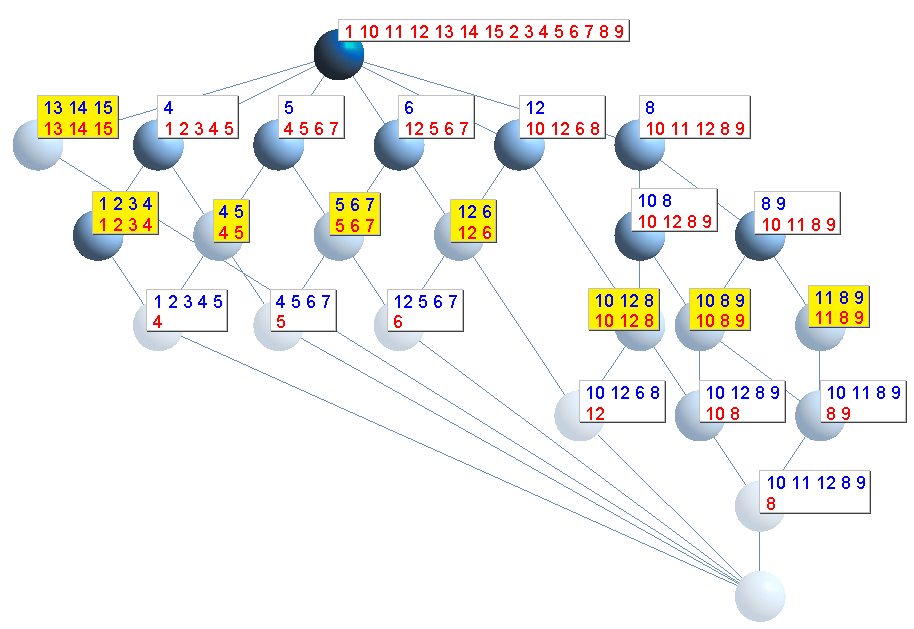}  
      \caption{The concept lattice $\mathcal{L}(\Tilde{\mathbb{K}})$.}
      \label{toylattice}
\end{figure}

We then build the concept lattice $\Tilde{\mathbb{K}}$
of our example as shown in Figure~\ref{toylattice}.
\subsection{Identifying Interesting cliques and Irrelevant bridges}
With a little more analysis of the constructed lattice $\mathcal{L}(\Tilde{\mathbb{K}})$, it is possible to identify the concepts in which the intent is equal to the extent.

\begin{definition}[identical concept] A formal concept $c=(A,B)$, with extent $A$ and intent $B$, is called an \textit{identical concept} if $A = B$, i.e., its extent and intent are identical.
\end{definition}
We use $\Tilde{\mathcal{C}}$ to denote the set of all the \textit{identical concepts}. \cut{The function \FuncSty{ExtractIC($\mathcal{L}$)} in Algorithm~\ref{pre} summarizes the pseudo-code for obtaining the set of all identical concepts $\Tilde{\mathcal{C}}$ of a lattice $\mathcal{L}$. \cut{Algorithm~\ref{pre}The function} \FuncSty{ExtractIC($\mathcal{L}$)} has a time complexity of $O(|\mathcal{C}|)$. 
\begin{algorithm}[!htb]
\caption{Extracting identical concepts from concept lattice.}
\begin{flushleft}
\Function{\textbf{Function} \FuncSty{ExtractIC($\mathcal{L}$)}}

\KwIn{Set of all concepts $\mathcal{C}$ in Lattice $\mathcal{L}$.}
\KwOut{Set of all identical concepts $\Tilde{\mathcal{C}}$.}
\end{flushleft}
  \begin{algorithmic}[1]    
      \State $\Tilde{\mathcal{C}} \leftarrow \emptyset$
      \For{\text{each concept} $c_i=(A_i,B_i) \in
      \mathcal{C}$}
      \If{$A_i \equiv B_i$}  
       \State $\Tilde{\mathcal{C}} \leftarrow \Tilde{\mathcal{C}} \cup \{c_i\} $;
       \EndIf
       \EndFor
       \State $\text{\textbf{Return}} \; \Tilde{\mathcal{C}}$;
    \end{algorithmic} \label{pre}
\end{algorithm}
}
\begin{proposition}\label{pro1} Given a social graph $\mathcal{U}$ and its corresponding concept lattice $\mathcal{L}(\Tilde{\mathbb{K}})$, an identical concept $c=(A,B) \in \mathcal{L}$ with $A=B$ and $|A|=k > 2$, actually represents a $k$-clique $l= \{g_i : g_i \in A \}$ in $\mathcal{U}$. 
\begin{proof}
Since an identical concept is a maximal square in the formal context, it represents a unit square matrix of size $k$ - as a sub-matrix of the modified adjacency matrix - and hence a $k$-clique.
Suppose now that $l = \{ g_i\}_{i=1}^k$ is a $k$-clique of $\mathcal{U}$ with $k>2$. Then, from Definition~\ref{Cl}, for any two object nodes $g_i, g_j$ in $l$, there exists an edge $(g_i, g_j)$ in $\mathcal{U}$ that connects the two objects. Using Eq.~\eqref{Fcontext}, the constructed $k \times k$ modified adjacency matrix  $\Tilde{\mathbb{K}}(l, l,\mathcal{I}_{l})$ that expresses the clique $l$ clearly defines a matrix consisting of all 1s. Such a matrix coincides with the identical concept $\Tilde{c}=(\{g_i\}_{i=1}^k,\{g_i\}_{i=1}^k)$ in which both extent $A$ and intent $B$ contain only the object nodes of $l$. This implies that a $k$-clique whose node set is $l = \{ g_i : g_i \in A \}$ is equivalent to an identical concept $\Tilde{c}=(A,B)$ such that $A=B=\{g_i\}_{i=1}^k$.
\end{proof}
\end{proposition}
For example, the identical concept $c=(\{5,6,7\},\{5,6,7\})$ of the concept lattice in Figure~\ref{toylattice} captures the 3-clique with nodes $l=\{5,6,7\}$ of the network in Figure~\ref{toygraph}. 

\begin{figure*}[!htbp]
  \begin{center}
  \centering
      \includegraphics[height=2.3in,width=2.9in]{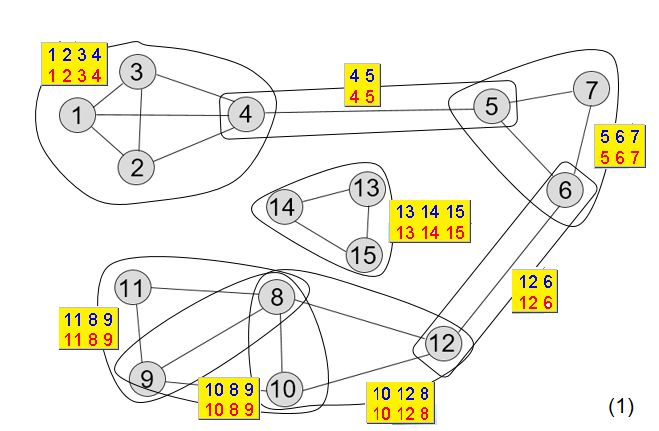}
      \vspace{6pt}
      \includegraphics[height=2.3in,width=2.9in]{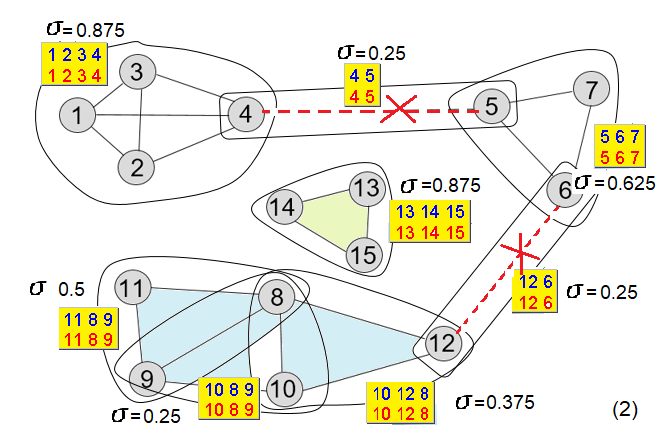}
      \includegraphics[height=2.3in,width=2.9in]{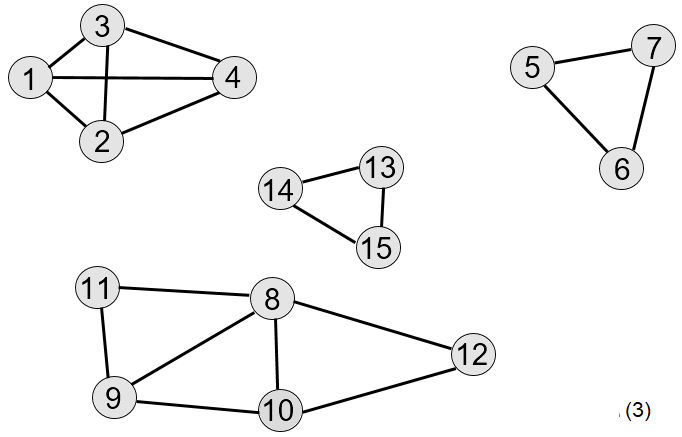}
    \end{center}
      \caption{How COIN algorithm works on the illustrative example. (1) Extracting the identical concepts that represent cliques and bridges. (2) Using the approximated stability index of identical concepts to cut noisy bridges, \textit{e.g.}, $\{(4,5),(6,12)\}$, and detect isolated maximum cliques, \textit{e.g.}, $\{13,14,15\}$. (3) Percolating the remaining relevant cliques, \textit{e.g.}, $\{11,8,9\}, \{10,8,9\}$ and $\{12,10,8\}$ to get the final predicted communities.}
      \label{toycom}
\end{figure*}

Therefore, from Proposition~\ref{pro1}, we can extract the cliques of the network $\mathcal{U}$ by identifying their corresponding identical concepts in $\mathcal{L}$. For instance, the identical concepts  $\Tilde{\mathcal{C}}$ appear in 'yellow' in Figure~\ref{toylattice}, and represent all the cliques and bridges as shown in Figure \ref{toycom}-1. Now, the role of concept interestingness comes into play. That is, we can measure how much noise exists in a clique by computing the stability index of its corresponding identical concept. The noise of a clique indicates how its objects are cohesive to each other and separable from other objects in the graph. Now, the question is how clique cohesion (looking for tightly cohesive groups) and separation (seeking for highly separated cliques) are inherited from the noise of their corresponding identical concepts. At a high level, this could be illustrated as follows. First, the stability captures the noise of an identical concept by estimating how its objects depend on the removal of each individual object. In fact, this measure of noisiness quantifies the connectivity among these objects in the corresponding clique. That is, much noise in the clique means that many objects are not cohesive and need to be removed to disconnect the clique. Second, the stability estimates the specificity of the object-object links of the identical concept with respect to the one-mode formal context. Thus, it assesses how these objects, in the corresponding clique, are influenced by the ties that hold between each individual object and other objects in the graph. That is, it approximately quantifies how these objects are strongly connected with other objects outside the clique. We call a clique \textit{relevant} if it contains a very small amount of noise, i.e., its objects have a high cohesion and and a low separability.

Consequently, the stability value of an identical concept approximates the probability that its corresponding clique is a portion of a potential community. For instance, if an identical concept has the highest stability value, then the involved objects of its corresponding clique are highly cohesive and completely separable from other objects in the graph. This, in fact, renders such a clique an isolated maximum one that likely forms a stand-alone community. On the contrary, an identical concept of size $2$, which has a low stability value, could identify a very noisy 2-clique (and hence, a non-trivial bridge in the graph) that is probably not a part of any potential community.                         
\begin{proposition}\label{pro2} Given a social graph $\mathcal{U}$ and its corresponding concept lattice $\mathcal{L}(\Tilde{\mathbb{K}})$, an identical concept $\Tilde{c}=(A,B) \in \mathcal{L}$ with $A=B$ and $|A|=k > 2$, represents a corresponding isolated maximum k-clique $l = \{ g_i : g_i \in A \}$ in $\mathcal{U}$ where $\Tilde{c}$ has the highest value of the stability index:
\begin{equation}
 \sigma(\Tilde{c}) = \frac{2^{|A|}-1}{2^{|A|}}    
\end{equation}
\begin{proof}
The proposition is held once we prove that: (1) $l = \{ g_i\}_{i=1}^k$ is represented by an identical concept; (2) This identical concept has the highest stability index.

(1) Suppose that $l = \{ g_i\}_{i=1}^k$ is an isolated maximum clique of size $k$ in $\mathcal{U}$. Since every ``isolated maximum'' $k$-clique has all the properties of a $k$-clique, then, from Proposition~\ref{pro1}, it can be easily demonstrated that $l$ has a modified adjacency matrix that defines a $k \times k$ all-ones matrix, and therefore $l$ is equivalent to an identical concept $\Tilde{c}=(A,B)$ such that $A=B=\{g_i\}_{i=1}^k$. 

(2) From Definitions~\ref{MC} and~\ref{IMC} of a maximal and isolated clique, we know that there is no edge that connects any object $g_i$ inside $l$ to any other object $g_a \in \mathcal{G} \setminus l$ outside $l$. Thus, the all-ones matrix of $l$ defines a sub-matrix of the whole one-mode formal context $\Tilde{\mathbb{K}}$, in which all elements in $\Tilde{\mathbb{K}}$ that define the relations among the objects $\{g_i\}_{i=1}^k$ outside $l$, are zeros. 
From the definition of unit matrix that defines $l$, we have:
    \begin{equation}
        \forall e \in \mathcal{P}(A), e \neq \emptyset \Rightarrow  e^{\prime{}} = A = B 
    \end{equation}
That is, except the empty set, all the other elements of the powerset $\mathcal{P}(A)$ satisfy the stability condition in the numerator of Eq.~\eqref{stability1}. This implies that the numerator of Eq.~\eqref{stability1}, in the stability of $\Tilde{c}$, is equal to the size of the powerset after excluding only the empty set. Thus, we have:
    \begin{equation}
        \sigma(\Tilde{c}) =  \frac{|\mathcal{P}(A)|-1}{|\mathcal{P}(A)|} = \frac{2^{|A|}-1}{2^{|A|}}
    \end{equation}
This implies that the stability of the identical concept, $\Tilde{c}=(A,B)$ with  $A=B=\{g_i\}_{i=1}^k$, is equal to $\frac{2^{|A|}-1}{2^{|A|}}$, and hence increases with the size of its corresponding $k$-clique. 
\end{proof}
\cut{\begin{proof}
see Appendix~\ref{p2}
\end{proof}}
\end{proposition}

For example, the identical concept $c=(\{13,14,15\},\{13,14,15\})$ in Figure~\ref{toylattice} captures the isolated maximum 3-clique $l=\{13,14,15\}$ shown in Figure~\ref{toygraph}, and has its highest stability value $\sigma(\Tilde{c}) = \frac{2^{3}-1}{2^{3}}=0.875$. 

\begin{proposition}\label{pro3} Given a social graph $\mathcal{U} $ and its corresponding concept lattice $\mathcal{L}(\Tilde{\mathbb{K}})$, an identical concept $c=(A,B) \in \mathcal{L}$, with $A=B= \{g_i,g_j\}, \; |A|= 2$, represents a corresponding non-trivial bridge $(g_i,g_j)$ in $\mathcal{U}$, and $\Tilde{c}$ has the
following stability index:
\begin{equation}
 \sigma(\Tilde{c}) = \frac{1}{4}    
\end{equation}
\begin{proof}
The proposition is held once we prove that: (1) a bridge is represented by an identical concept with an extent and intent involving only the two objects of the bridge; (2) This identical concept has a stability value of $\frac{1}{4}$.

(1) Let $b=(g_i,g_j)$ be a non-trivial bridge between two components $\mathcal{T}_i$ and $\mathcal{T}_j$ of $\mathcal{U}$ such that $g_i \in \mathcal{T}_i$ and $g_j \in \mathcal{T}_j$. From Eq.~\eqref{Fcontext}, the $2 \times 2$ modified adjacency matrix of $b$ defines a unit matrix $J_b$. Now, since each object of the bridge $b$ belongs to a different component, then its $J_b$ matrix is also a sub-matrix of the whole one-mode formal context $\Tilde{\mathbb{K}}$ such that we have the following two properties:
\cut{\begin{enumerate}[(i)]
\item Zero elements in $\Tilde{\mathbb{K}}$ between the object $g_i$ and $\mathcal{T}_j \setminus \{g_j\}$. This means that all objects of the component $\mathcal{T}_j$ except the object $\{g_j\}$ have a zero value
\item Zero elements in $\Tilde{\mathbb{K}}$ between the object $g_j$ and $\mathcal{T}_i \setminus \{g_i\}$, i.e., all objects of the component $\mathcal{T}_i$ except the object $\{g_i\}$ have a zero value.
\end{enumerate}}

\begin{enumerate}[(i)]
\item  $(g_i, g_p) = 0$ $\forall g_i \in \mathcal{T}_i$ and  $g_p \in \mathcal{T}_j \setminus \{g_j\}$ 
\item $(g_j, g_p) = 0$ $\forall g_j \in \mathcal{T}_j$ and  $g_p \in \mathcal{T}_i \setminus \{g_i\}$
\end{enumerate}

The modified adjacency matrix $J_b$ of the bridge can be used to extract, from $\Tilde{\mathbb{K}}$, an identical concept $\Tilde{c}=(\{g_i,g_j\},\{g_i,g_j\})$ where both its intent and extent contain the two objects (nodes) of the bridge.  

(2) $\mathcal{P}(\Tilde{c}) = \{\emptyset,\{g_i\},\{g_j\},\{g_i,g_j\}\}$ is the powerset of the identical concept $\Tilde{c}=(\{g_i,g_j\},\{g_i,g_j\})$. Based on the definition and properties of the modified adjacency matrix $J_b$ of the bridge, only one subset $\{g_i,g_j\} \in \mathcal{P}(\{g_i,g_j\})$ satisfies the stability condition in the numerator of Eq.~\eqref{stability1}, while the other subsets, i.e.,$\{\emptyset,\{g_i\},\{g_j\}\}$ do not. Thus, the numerator of Eq.~\eqref{stability1} contains only one subset. This implies that the stability of the identical concept $\Tilde{c}=(\{g_i,g_j\},\{g_i,g_j\})$ is equal to $\frac{1}{2^{|\{g_i,g_j\}|}} = \frac{1}{2^{|2|}} = \frac{1}{4}$.
\end{proof}
\cut{\begin{proof}
see Appendix~\ref{p3}
\end{proof}}
\end{proposition}
For example, in Figure~\ref{toylattice}, the identical concept $c=(\{4,5\},\{4,5\})$ has a stability $\sigma(\Tilde{c}) = 0.25$ and captures the non-trivial bridge $(4,5)$ in the graph of Figure~\ref{toygraph}. 
Object $4$ belongs to $\mathcal{T}_i=\{1, 2, 3, 4\}$ while object $5$ is an element of $\mathcal{T}_j=\{5, 6, 7\}$.

\subsection{COIN Algorithm For Detecting Communities}
\begin{algorithm}[!htbp]
\caption{The COIN algorithm for detecting communities.}
\begin{flushleft}
\textbf{Input:} Set of all identical concepts $\Tilde{\mathcal{C}}$ in the concept lattice $\mathcal{L}(\mathbb{K})$. \\ 
\textbf{Output:} Set of all communities $\mathcal{D}$ in network $\mathcal{U}$.
\end{flushleft}
 \begin{algorithmic}[1]    
      \State $\mathcal{D} \leftarrow \mathcal{R} \leftarrow \emptyset$
      \Statex \text{\texttt{//\cut{ Stage 1:} Extract isolated cliques and remove bridges}}
      \For{\text{each concept} $\Tilde{c}_i=(A_i,B_i) \in \Tilde{\mathcal{C}}$}
      \State $\sigma(\Tilde{c}_i) \leftarrow \text{Approximate the stability of $\Tilde{c}_i$ using LDS}$
      \If{$\sigma(\Tilde{c}_i) = \frac{2^{|A_i|}-1}{2^{|A_i|}}$}
      \Statex \text{\texttt{// $\Tilde{c}_i$ is an isolated maximum clique}}
      \State $\mathcal{D} \leftarrow \mathcal{D} \cup \{\Tilde{c}_i\}$  \text{\texttt{  // $\Tilde{c}_i$ as a community}}
      \State $\Tilde{\mathcal{C}} \leftarrow \Tilde{\mathcal{C}} \setminus \{\Tilde{c}_i\}$
      \EndIf
       \If{$|A_i| = 2 \text{ and } \sigma(\Tilde{c}_i)  = \frac{1}{4}$}
       \Statex \text{\texttt{// Cut $\Tilde{c}_i$, which is a noisy non-trivial bridge}}
       \State $\Tilde{\mathcal{C}} \leftarrow \Tilde{\mathcal{C}} \setminus \{\Tilde{c}_i\}$
       \EndIf
     \EndFor
     \Statex \text{\texttt{//\cut{ Stage 2:} Percolate adjacent relevant cliques}}
       \For{$\Tilde{c}_i=(A_i,B_i),\Tilde{c}_j=(A_j,B_j) \in \Tilde{\mathcal{C}}$}
       \State $a_{ij} \leftarrow \min(|A_i|,|A_j|)$
       \If{$|A_i \cap A_j| \geqslant a_{ij}-1$}
       \State $\Tilde{c}_{ij} \leftarrow \text{Merge}(\Tilde{c}_i,\Tilde{c}_j)$
       \State $\Tilde{\mathcal{C}} \leftarrow \Tilde{\mathcal{C}} \setminus \{\Tilde{c}_i,\Tilde{c}_j\}$
       \State $\Tilde{\mathcal{C}} \leftarrow \Tilde{\mathcal{C}} \cup \{\Tilde{c}_{ij}\}$
       \EndIf
       \EndFor
      \State $\mathcal{D} \leftarrow \mathcal{D} \cup \Tilde{\mathcal{C}}$
       \State \text{\textbf{return}}  $\mathcal{D}$
   \end{algorithmic} \label{coinalgo}
\end{algorithm}

Algorithm \ref{coinalgo} gives the pseudo-code for the COIN community detection algorithm. For clarity, Figure \ref{toycom} shows the steps of COIN applied to the example in Figure \ref{toygraph}. The algorithm takes as input the set of all identical concepts $\Tilde{\mathcal{C}}$ that capture all the cliques and bridges.
\cut{, where $\Tilde{\mathcal{C}}$ can be obtained by calling the function in Algorithm~\ref{pre}. }
The COIN algorithm then goes through two stages. At the first stage, as shown in Figure \ref{toycom}-2, it uses an efficient Low-discrepancy sampling (LDS) method in \cite{IbrahimMRokia18} to approximate the stability of each identical concept (line 3). Then, it distinguishes two types of identical concepts based on the estimated stability value. The first type is an identical concept that has its highest stability value. From Proposition \ref{pro2}, such identical concept represents an isolated community, and we therefore detect it as a community and move it into the set of final communities $\mathcal{D}$ (lines 4-7). The second type is an identical concept that contains only two objects and has a stability value of $\frac{1}{4}$ (line 8). According to Proposition \ref{pro3}, this identical concept represents a noisy non-trivial bridge between two potential communities. Thus, we cut this bridge by removing it from the set $\Tilde{\mathcal{C}}$ (line 9). At this stage, $\Tilde{\mathcal{C}}$ contains only the subset of identical concepts which have a stability higher than $\frac{1}{4}$. Such concepts capture the relevant cliques.      

At the second stage, the algorithm iteratively applies a pairwise percolation of every two neighboring relevant cliques $\Tilde{c}_i,\Tilde{c}_j \in \Tilde{\mathcal{C}}$ if they share at least $a_{ij}-1$ common objects (lines 12-19), where $a_{ij}$ is the smallest number of objects in the extents of $\Tilde{c}_i$ and $\Tilde{c}_j$. Finally, the algorithm moves the components, obtained after the completion of percolation, into $\mathcal{D}$ (line 20), and it then returns the final set of detected communities as shown in Figure \ref{toycom}-3. 

Assume now that the network contains an additional component that represents the following star graph:
$$G_5 = ( \{16, 17, 18, 19\}, \{(16, 17), (16, 18), (16, 19)\})$$ 

Then, three $2$-cliques (equivalent to identical concepts) are identified: $(16, 17), (16, 18)$ and $(16, 19)$ with a stability equal to $0.5 > 0.25$. Since they do not represent noisy non-trivial bridges, they will be merged at the second step of the algorithm to appear as a (star) community. 

\textbf{Complexity analysis}. The first-stage of COIN has a time complexity of  $O(|\Tilde{\mathcal{C}}| \times \xi)$, while the second-stage has $O(|\Tilde{\mathcal{C}}|^2)$ time complexity, where $\Tilde{\mathcal{C}}$ is a set of the identical concepts and $\xi$ is the time needed to approximate the stability index of an identical concept. Thus, the total time complexity of COIN algorithm is $O(|\Tilde{\mathcal{C}}| \times \xi + |\Tilde{\mathcal{C}}|^2)$.

\section{experimental Evaluation}
\label{Tests}
The main goal of our experimental evaluation is to investigate
the following key questions:
\begin{itemize}
    \item (Q1) Is COIN more accurate than the state-of-the-art community detection algorithms?
    \item (Q2) Is COIN scalable compared other prominent algorithms for detecting communities in social networks?
\end{itemize}
\subsection{Methodology}
We started our experiments by first selecting four datasets of real-life social networks:
\begin{enumerate}
    \item \textit{Zachary's karate club} (Karate) \cite{zachary1977information} which describes $34$ members of the karate club, showing $78$ pairwise connection between members who interacted outside the club. A conflict arose between the administrator (member No. 0) and the instructor (member No. 33), which led to the scission of the club into two non-overlapped communities around the two members $0$ and $33$; \item \textit{American College football} (Football) \cite{girvan2002community} which is a network that represents the schedule of games between college football teams in a single season
    \item \textit{Bottlenose Dolphin} (Dolphins) \cite{lusseau2003bottlenose} which describes a network of frequent associations between $62$ dolphins in a community living of Doubtful Sound in New Zealand
    \item \textit{Books about US politics} (PolBooks) \cite{krebs2008network} is a network of books about US politics published around the time of the 2004 presidential election and sold by the online bookseller Amazon.com.
\end{enumerate}
Table~\ref{data} briefly summarizes these datasets which are publicly available at the following URL\footnote{ \url{http://www-personal.umich.edu/~mejn/netdata/}.}. 
\begin{table}[!htbp]
\centering
\caption{A brief description of the tested social networks. $\mathcal{G}$ is the number of object nodes, $\mathcal{I}$ is the number of edges, and $\hat{\mathcal{D}}$ is the number of ground truth communities.}
\label{data}
\begin{tabular}{|l|c|c|c|l|}
\hline
Name & \textbf{$\mathcal{G}$} & \textbf{$\mathcal{I}$} & \textbf{$\hat{\mathcal{D}}$} & \textbf{Description of group nature} \\ \hline
\textbf{Karate} & 34 & 78 & 2 & Membership after the division \\ \hline
\textbf{Football} & 115 & 615 & 12 & Team scheduling \\ \hline
\textbf{Dolphins} & 62 & 159 & 2 & Group of male/female dolphins \\ \hline
\textbf{PolBooks} & 105 & 441 & 3 & Group of books about US politics \\ \hline
\end{tabular}
\end{table}

To get answers to the proposed questions, we empirically evaluate our proposed COIN algorithm by comparing its results with the following state-of-the art community detection algorithms: 
\begin{enumerate}
    \item \textbf{Louvain} \cite{blondel2008fast} which is an algorithm that uses a heuristic that maximizes the modularity 
    \item Clique percolation method \textbf{CPM} \cite{palla2005uncovering,evans2010clique,adamcsek2006cfinder} which builds up the communities by percolating adjacent k-cliques, where two $k$-cliques are considered adjacent if they share $k-1$ object nodes 
    \item  Girvan-Newman \textbf{GN} (also called Edge-betweenness) \cite{newman2004finding,girvan2002community} which is a hierarchical decomposition algorithm that progressively removes edges based on the decreasing order of their betweenness scores 
\cut{    \item Label propagation \textbf{LP} \cite{raghavan2007near}: iteratively re-assign every object node the most frequent (the majority) label of its neighbors in a synchronous manner. It is very fast but non-deterministic, so it is highly sensitive to the initial parameters}
    \item \textbf{WalkTrap} \cite{pons2005computing} which performs random walks to get the walks that are more likely to stay within the same communities while only a few edges are outside a given community\cut{ --- serves as a good baseline here }
    \item Fast greedy modularity \textbf{FGM} \cite{clauset2004finding} which is the fastest greedy community analysis algorithm that detects the set of communities based on the optimization of the modularity score
    \item \textbf{InfoMAP} \cite{rosvall2008maps,rosvall2009map}: performs random walks to analyze the information flow through a network, and detects the community set that minimizes the expected description length of a random walker trajectory.\cut{ This algorithm starts with encoding the network into modules in a way that maximizes the amount of information about the original network. Then it sends the signal to a decoder through a channel with limited capacity. The decoder tries to decode the message and to construct a set of possible candidates for the original graph. The smaller the number of candidates, the more information about the original network has been transferred}
\end{enumerate}

We then assess the accuracy and scalability of results by recording the \textbf{average elapsed time} ($\tau$), and then calculate the following \textbf{Normalized mutual information} (NMI) metric \cite{danon2005comparing}:
\begin{equation}
    NMI(\mathcal{D},\hat{\mathcal{D}}) = \frac{-2 \sum_{i=1}^{|\hat{\mathcal{D}}|} \sum_{j=1}^{|\mathcal{D}|} n_{ij} \log{(\frac{n_{ij}  n}{n_i  n_j}})}{\sum_{i=1}^{|\hat{\mathcal{D}}|} n_{i} \log{(\frac{n_{i}}{n}}) + \sum_{j=1}^{|\mathcal{D}|} n_{j} \log{(\frac{n_{j}}{n}})}
\end{equation}
Where $n = |\mathcal{G}|$ is the number of object nodes. $\mathcal{N}$ is a confusion matrix where the rows correspond to \textit{ground-truth communities} $\hat{\mathcal{D}}$ and the columns correspond to \textit{predicted communities} $\mathcal{D}$ found by a given community detection algorithm. Each element $n_{ij} \in \mathcal{N}$ is the number of object nodes in the \textit{i-th} ground-truth community that appear in the \textit{j-th} predicted community. $n_i = \sum_j n_{ij}$ is the sum over row $i$ of $\mathcal{N}$, and $n_j = \sum_i n_{ij}$ is the sum over column $j$ \footnote{NMI metric $\in [0,1]$. That is, if the predicted communities $\mathcal{D}$ are identical to the ground-truth ones $\hat{\mathcal{D}}$, then NMI metric is equal to $1$. On the contrary, if they are totally independent, then NMI metric is $0$.}.

Furthermore, to guarantee a fair comparison, we conducted our empirical study with the chosen algorithms under the following setting:
\begin{itemize}
    \item Re-run all tested algorithms $100$ times, and report the average result \cut{\textit{This could be important to fairly judge the performance of non-deterministic algorithms such as \textbf{WalkTrap}, which it is sensitive to the initial parameters (e.g., initial labeling) and often yields different result at each run.}} 
    \item Vary the parameter $k \in \{3,4,5\}$ in the k-cliques percolation process of \textbf{CPM}, and report the average result
    \item Use \textbf{WalkTrap} to serve as a good baseline for assessing the accuracy and scalability
    \item Tune the resolution parameter $\eta$ of \textbf{Louvain} and \textbf{FGM} algorithms in the range $(0,2]$, and record the best results. Traditionally, $\eta$ is assigned a value $1$ by default. But one can set $\eta$ to a value greater than $1$ to find a larger number of smaller communities, and assign it a value less than $1$ to detect a smaller number of larger communities.
\end{itemize}

All the experiments were run on an Intel(R) Core(TM) i7-2600 CPU @ 3.20GHz computer with 16GB of memory under Windows $10$. We implemented our COIN algorithm as an extension to the FCA Concepts 0.7.11 package that is implemented by Sebastian Bank.\footnote{Publicly available at: \url{https://pypi.python.org/pypi/concepts}.}

\subsection{Results}\label{result}
\begin{table}[!htbp]
\centering
\caption{NMI-score of the considered community detection algorithms  on the tested social networks. The value between (.) is the number of predicted communities.}
\label{NMIscore}
\begin{tabular}{|l|l|l|l|l|}
\hline
\multicolumn{1}{|c|}{\multirow{2}{*}{\textit{Methods}}} & \multicolumn{4}{c|}{\textit{Social networks}} \\ \cline{2-5} 
\multicolumn{1}{|c|}{} & \textbf{Karate} & \textbf{Football} & \textbf{Dolphins} & \textbf{PolBooks} \\ \hline
\textbf{Louvain} & $0.597$ $(4)$ & $0.775$ $(10)$ & $0.526$ $(5)$ & $0.591$ $(4)$ \\ \hline
\textbf{CPM} & $0.294$ $(3)$ & $0.903$ $(13)$ & $0.38$ $(4)$ & $0.455$ $(6)$ \\ \hline
\textbf{COIN} & \textbf{0.837} $(2)$ & \textbf{0.978} $(12)$ & \textbf{1.00} $(2)$ & \textbf{0.885} $(3)$ \\ \hline
\textbf{GN} & $0.789$ $(2)$ & $0.807$ $(10)$ & $0.890$ $(2)$ & $0.543$ $(5)$ \\ \hline
\cut{\textbf{LP} & $0.342$ $(2)$ & $0.841$ $(10)$ & $0.554$ $(4)$ & $0.693$ $(3)$ \\ \hline}
\textbf{WalkTrap} & $0.546$ $(5)$ & $0.363$ $(10)$ & $0.465$ $(8)$ & $0.569$ $(4)$ \\ \hline
\textbf{FGM} & $0.706$ (3) &  $0.513$ $(6)$ & $0.454$ $(4)$ & $0.568$ $(4)$ \\ \hline
\textbf{InfoMAP} & $0.712$ $(3)$ & $0.890$ $(12)$ & $0.557$ $(6)$ & $0.567$ $(6)$ \\ \hline
\end{tabular}
\end{table}

\begin{figure}[!htb]
  \begin{center}
      \includegraphics[height=2.0in,width=3.2in]{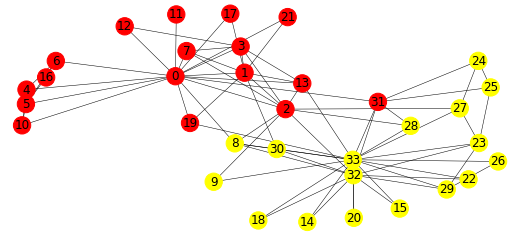} \\
    \end{center}
      \caption{The predicted communities of \textit{Karate} network obtained by COIN algorithm.}
      \label{result1}
\end{figure}

\begin{figure}[!htb]
  \begin{center}
      \includegraphics[height=2.9in,width=3.6in]{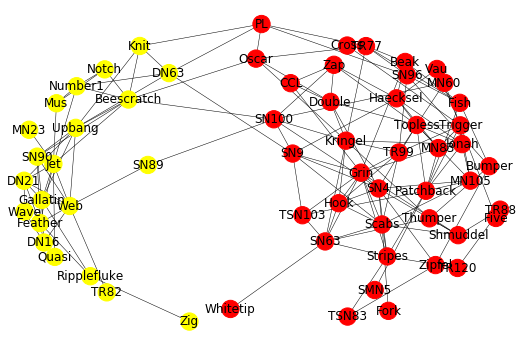} \\
    \end{center}
      \caption{The predicted communities of \textit{Dolphin} network obtained by COIN algorithm.}
      \label{result2}
\end{figure}

\begin{figure*}[!htb]
  \begin{center}
      \includegraphics[height=6.0in,width=6.0in]{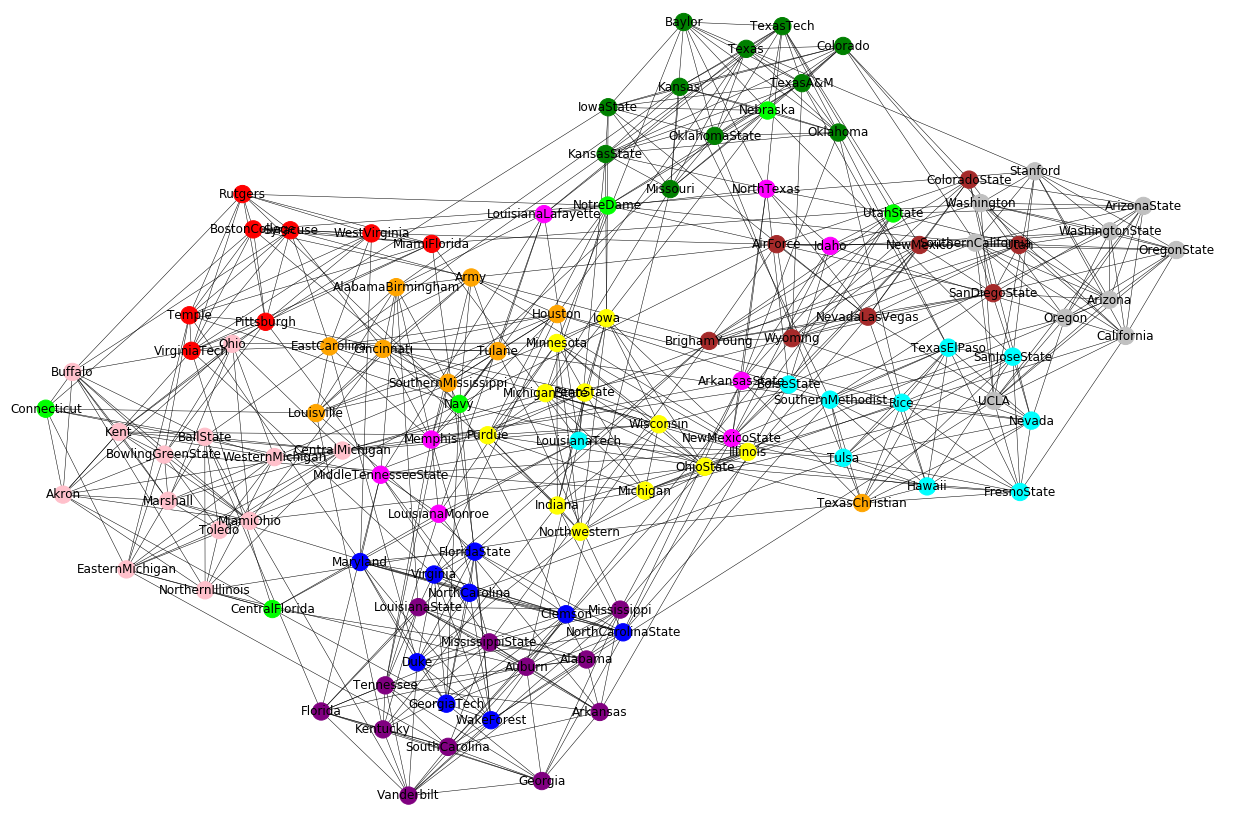} \\
    \end{center}
      \caption{The predicted communities of \textit{Football} network obtained by COIN algorithm.}
      \label{result3}
\end{figure*}

\begin{figure*}[!htb]
  \begin{center}
      \includegraphics[height=6.0in,width=6.0in]{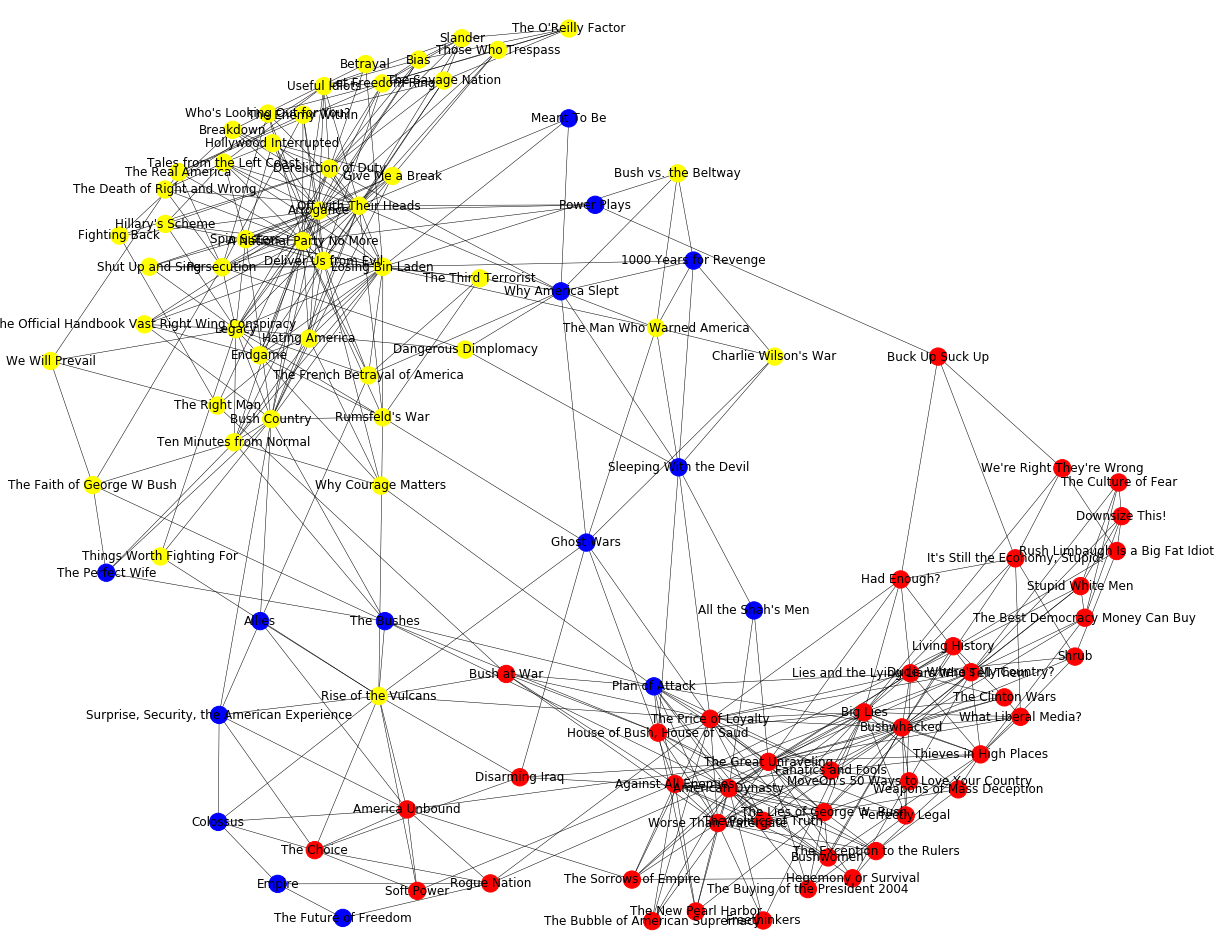} \\
    \end{center}
      \caption{The predicted communities of \textit{PolBooks} network obtained by COIN algorithm.}
      \label{result4}
\end{figure*}

In terms of accuracy, the results in Table~\ref{NMIscore} illustrate that COIN is the most accurate of all the compared algorithms, achieving the best NMI score on the four tested social networks. On the Karate dataset, it produces two communities that are approximately $84\%$ correct compared to the two ground-truth ones, and in which only the object node $31$ is wrongly detected in the `red' community instead of the `yellow' one as shown in Figure~\ref{result1}. For the Dolphins dataset, COIN detects the two communities that are identical to the ground truth ones as shown in Figure~\ref{result2}. On the Football dataset, COIN predicts the $12$ communities that are approximately $98\%$ correct, where only two object nodes (`nebraska') and (`memphis') are wrongly detected in `lime' and `magenta' communities instead of `green' and `orange' true communities respectively (see Figure~\ref{result3}). It also obtains the three communities of Polbooks dataset with $89\%$ accuracy, where as shown in Figure~\ref{result4}, the object nodes `allies', `the bushes' and `bush at war' predicted to be in `blue' and `red' communities instead of the `yellow' true one. 

InfoMAP and CPM come close behind COIN on Fooball, but considerably further behind on both Dolphins and Polbooks. On  both Karate and PolBooks, WalkTrap was marginally less accurate than Louvain, but more accurate than CPM. The latter outperforms both FGM and Louvain on the Football dataset. Remarkably, CPM has poor results on both the Karate and Dolphins datasets, and both InfoMAP and Louvain clearly outperform FGM on the Football dataset.

\begin{figure}[!htb]
  \begin{center}
      \includegraphics[height=2.2in,width=3.1in]{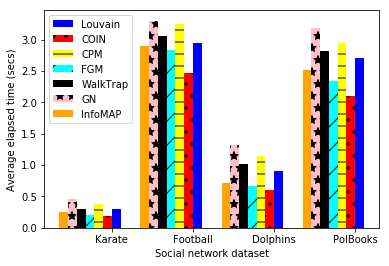} \\
    \end{center}
      \caption{The average elapsed time $\tau$ of the underlying community detection algorithms on the tested social networks.}
      \label{resulttime}
\end{figure}

In terms of computational time, the results in Figure~\ref{resulttime} show that COIN is relatively faster than FGM, which is the quickest community detection algorithm among all other tested ones: COIN needs only an average time of $0.186$, $2.46$, $0.61$ and $2.1$ seconds to detect communities on the four datasets respectively, while FGM takes $0.21$, $0.2.92$, $0.72$, and $2.34$ seconds. InfoMAP comes close behind FGM on all datasets, and both of them are considerably fast compared to the baseline WalkTrap on all underlying social networks. Aside from FGM algorithm, COIN clearly prevails over both CPM, Louvain and WalkTrap on all datasets by a significant  margin, while saving  time by more than twice compared to GN on Karate and Dolphin datasets. In practice, this is due to the fact that COIN mainly detects communities based on the set of identical-concepts $\Tilde{\mathcal{C}}$, which is frequently small compared to the sets of objects and edges that are used by all other tested community detection algorithms.  

Apart from COIN, FGM and InfoMAP, none of the tested algorithms --- the CPM, WalkTrap, and Louvain --- shows its clear superiority over the others. Remarkably, Louvain is very competitive with WalkTrap, and CPM slightly competes with GN algorithm on all datasets.

Overall, the results in Figures~\ref{resulttime}-\ref{result4} and Table~\ref{NMIscore}, clearly show that COIN can detect communities more accurately than existing prominent algorithms  and even with less computational time.

\section{Conclusion}
\label{Conc}
We have proposed COIN, a two-stage method that exploits Formal Concept Analysis and concept stability to efficiently detect communities in one-mode social networks. All the \emph{identical concepts} that capture cliques and bridges are first extracted from the concept lattice. Then, the stability index of these concepts is used to identify relevant cliques and cut irrelevant bridges between communities. Finally, cliques with particular features are merged to obtain the final communities. Our method has been tested on different networks with different shapes of components (e.g., star, fully or strongly connected components, chains) and led to accurate results.
The empirical study on real-life social networks (see Section \ref{Tests}) shows that COIN can detect communities in a more accurate and efficient manner than other state-of-the-art methods.

In the future we plan to generalize COIN to detect \textit{overlapping} communities both in \textit{one-mode}, \textit{two-mode} data and even multi-layer networks. We also intend to propose an online (incremental) variant of COIN algorithm to tackle dynamic social networks and capture the evolution of their communities over time. Finally, we intend to increase the accuracy of COIN by exploiting other concept interestingness measures (e.g. separation index and concept probability \cite{Kuznetsov2015}) and intensively investigate the efficiency of COIN on big and dense social data networks.      



\bibliographystyle{unsrt}  
\bibliography{References}  

\end{document}